\documentclass[prl,twocolumn, superscriptaddress, a4paper, showpacs]{revtex4}
\usepackage{amsmath,amsfonts,amssymb}
\usepackage{graphicx}

\begin{document}

\title{Generalized superstatistics of nonequilibrium Markovian systems}
\author{Ihor Lubashevsky}
    \email{ialub@fpl.gpi.ru}
    \affiliation{Theory Department, A.M. Prokhorov General Physics Institute, Russian
    Academy of Sciences, \\ Vavilov Str. 38, 119991 Moscow, Russia}
    \affiliation{Institut f\"ur Physikalische Chemie,
    Westf\"alische Wilhelms Universit\"at M\"unster,  Corrensstr. 30,
    48149 M\"unster, Germany}
\author{Rudolf Friedrich}
    \email{fiddir@uni-muenster.de}
    \affiliation{Institut f\"ur Theoretische Physik,
    Westf\"alische Wilhelms-Universit\"at M\"unster, Wilhelm-Klemm-Str. 9, D-48149 M\"unster,
    Germany}
\author{Andrey Ushakov}
    \email{ushakov-an1986@rambler.ru}
    \affiliation{Faculty of Radioengineering Systems, Moscow Technical University of
    Radioengineering, Electronics, and Automation,
        \\ Vernadsky pros., 78, 117454 Moscow, Russia}
\author{Andreas Heuer}
    \email{andheuer@uni-muenster.de}
    \affiliation{Institut f\"ur Physikalische Chemie, Westf\"alische Wilhelms
    Universit\"at M\"unster,  Corrensstr. 30,
    48149 M\"unster, Germany}
    \affiliation{Center of Nonlinear Science CeNoS,
    Westf\"alische Wilhelms Universit\"at M\"unster, 48149 M\"unster, Germany}
\date{\today}
%
\begin{abstract}
The paper is devoted to the \textit{construction} of the superstatistical
description for nonequilibrium Markovian systems. It is based on Kirchhoff's
diagram technique and the assumption on the system under consideration to
possess a wide variety of cycles with vanishing probability fluxes. The latter
feature enables us to introduce equivalence classes called channels within
which detailed balance holds individually. Then stationary probability as well
as flux distributions are represented as some sums over the channels. The
latter construction actually forms the superstatistical description, which,
however, deals with a certain superposition of equilibrium subsystems rather
then is a formal expansion of the nonequilibrium steady state distribution into
terms of the Boltzmann type.
\end{abstract}

\pacs{02.50.Cw, 02.50.Ey, 02.50.Ga,  05.20.-y, 05.20.Dd, 05.70.Ln}

\maketitle

\section{Superstatistics and stationary states of nonequilibrium
Markovian systems}

During the last several years there has been considerably grown interest in the
description of essentially nonequilibrium systems using quasiequilibrium
notions, namely, the concept of ``superstatistics'' by Beck \& Cohen
\cite{SS1}. Briefly, it assumes the stationary state $\mathcal{P}_i^\text{st}$
of a nonequilibrium system can be written in the Boltzmann form with averaging
over possible fluctuations in the inverse temperature $\beta$,
\begin{equation}\label{1}
   \mathcal{P}_i^\text{st} = \int\limits_0^\infty d\beta\,
   f(\beta)\frac1{Z(\beta)}\exp\{-\beta E_i\}\,,
\end{equation}
where $f(\beta)$ is the probability distribution of the inverse temperature,
$E_i$ is the effective energy of the system state $i$, and $Z(\beta)$ is the
partition function for a fixed value of $\beta$. Representation~\eqref{1} is
actually a generalization of the so-called nonextensive statistics introduced
by Tsallis \cite{Tsallis} and in a integral form relates powerlike and
Boltzmann distributions \cite{PreSS}.

The main idea of superstatistics, however, seems to have a longer history
\cite{His1}, at least, expressions similar to Eq.~\eqref{1} can be found in
monograph by Lavenda \cite{His2} and this problem goes back to Szilard
\cite{His3} and Mandelbrot \cite{His3} as well as the results of Hungarian
school on information theory \cite{Hungary}. Moreover, in an effort to derive
representation~\eqref{1} starting from the general description of statistical
systems ones have met fundamental problems and inconsistencies
\cite{Prob1,Prob2,Prob3}. On the other hand, the superstatistical description
is rather natural especially for systems exhibiting large-scale fluctuations in
temperature gradients \cite{Sattin} or flow turbulence, where the
spatiotemporal fluctuations in the energy dispersion is the standard fact going
back to early work by Kolmogorov \cite{Kolmog}.

In spite of fundamental problems met in justifying expression~\eqref{1} the
number of papers dealing with the superstatistics has increased remarkably in
the last years. In particular, it has been applied to Lagrangian
\cite{Turb1,Turb2,Turb3,Turb31} and Eulerian turbulence
\cite{Turb4,Turb5,Turb71}, defect turbulence \cite{Turb8}, atmospheric
turbulence \cite{ATurb1,ATurb2}, cosmic ray statistics \cite{crs}, statistics
of solar flares \cite{solfl}, hadronization of quark matter \cite{hqm},
small-world networks \cite{swn}, multi-components self-gravitating systems and
collisionless stellar systems \cite{cgs}, transitions between regular-chaotic
dynamics \cite{amdg1,amdg2}, particle ensembles with fractional reactions
\cite{andiff}, analysis of time series \cite{timeseries}, econophysics
\cite{ecph1,ecph2,ecph3}.

In some sense expression~\eqref{1} can be interpreted in two fashions. The
first way is to regard expression~\eqref{1} as a rather formal expansion of the
stationary distribution for a nonequilibrium system over terms having the
Boltzmann form. In this case the main problem is the relationship between the
weights $f(\beta)$ of this expansion and specific random processes governing
the system dynamics, which is currently the main direction of researches
carried out in this field (see, e.g., Ref.~\cite{Beck1000}). The other could be
an attempt to represent a nonequilibrium system with nonzero stationary
probability flux as a certain superposition of its subsystems being
equilibrium, i.e. within which the detailed balance holds individually. Exactly
this idea is the goal of our study. The purpose of the present paper is to
implement this approach to describing the stationary properties of a
nonequilibrium Markovian system. Naturally, the final integral over possible
subsystems with local equilibrium has to be of a more general form than
formula~\eqref{1} and is reduced to it in special cases only, which is the
reason of using the term ``generalized superstatistics''.

There has been a great deal of studying nonequilibrium Markovian systems within
the frameworks of the master equation, for a review see, e.g.,
Refs.~\cite{Schank,NESS0,NESS00}. During the last decades many nonequilibrium
systems as well as systems of other nature where the equilibrium notion is
irrelevant came into view of physical society. This, in particular, has
reawaken the interest to the general steady state properties exhibited by
Markovian systems without detailed balance and posed a question about their
minimal mesoscopic description called dynamic equivalence classes
\cite{NESS1,NESS2} (see also Ref.~\cite{NESS01}). Within the latter approach it
makes an attempt to find aggregated characteristics determining the steady
state distribution as well as probability fluxes in a nonequilibrium system
without detailed description of all the transition rates between the system
states. In the present work we actually follow the spirit of this idea.

\section{Quasiequilibrium channels of a nonequilibrium Markovian system}

We consider a Markovian system with a finite number of states $\left\{
i\right\}$. This number, however, may take any large value, so, there should be
a feasibility to generalize the following constructions to systems with
infinite number of states. The system evolution is described by the master
equation
\begin{equation}
\frac{d\mathcal{P}_{i}}{dt}=\sum_{j\neq i}\left\langle i|j\right\rangle
\mathcal{P}_{j}-\left\langle j|i\right\rangle \mathcal{P}_{i}  \label{mp:1}
\end{equation}
written for the distribution function $\mathcal{P}_{i}$, where $\left\langle
i|j\right\rangle $ stands for the rate of system transitions to state $i$ from
state $j$. All the system states make up a graph $\mathbb{G}$ whose edges
present possible transitions between the states. Without loss of generality we
may adopt two assumptions about this graph, following, e.g.,
Ref.~\cite{Schank}. First, if there exists a transition from some state $j$ to
some state $i$, i.e. $\left\langle i|j\right\rangle >0$, then the reverse
transition is also possible, i.e. \thinspace $\left\langle j|i\right\rangle
>0$. The opposite case is included within the limit $\left\langle
j|i\right\rangle \rightarrow 0$. Second, the graph $\mathbb{G}$ is connected,
which implies that for each pair of states, i.e. graph nodes $(i,j)$, there
exists at least one path $\mathbb{P}_{ij}$ on the graph $\mathbb{G}$ (sequence
of joint edges) connecting them. Otherwise, the physical system behind the
graph $\mathbb{G}$ can be decomposed into two or more independent subsystems
analyzed individually.

\begin{figure}
\begin{center}
\includegraphics[width=0.7\columnwidth]{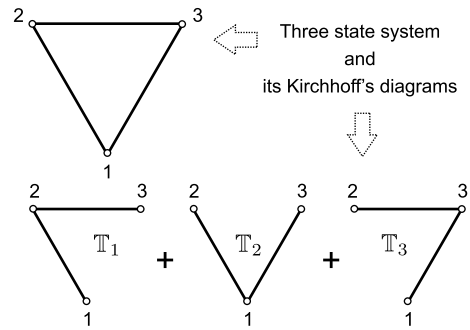}
\end{center}
\caption{A three state system and its division into maximal trees.}
\label{Fig0}
\end{figure}
\begin{figure}
\begin{center}
\includegraphics{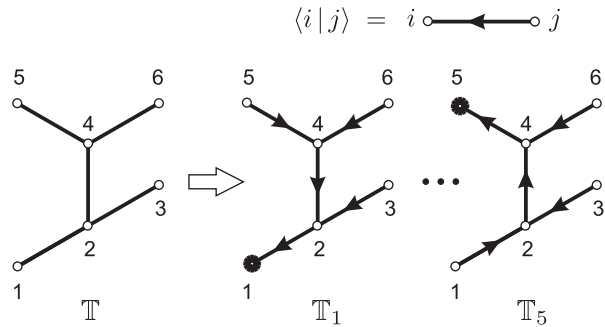}
\end{center}
\caption{Example of a maximal tree for a six state system and its realization
for two states, 1 and 5. The correspondence between the graph edges and the
transition rates $\left<i|j\right>$ is shown above.} \label{Fig1}
\end{figure}

\subsection*{Kirchhoff's diagrams}

The following part of the paper will deal only with the steady state properties
of such Markovian systems described by the stationary solution $
P_{i}=\mathcal{P}_{i}^{\text{st}}$ of the master equation~(\ref{mp:1}). This
solution is represented using Kirchhoff's diagram technique (see, e.g.
Refs.~\cite{Schank,Haken}). It applies to the notion of maximal trees. By
definition, a maximal tree $\mathbb{T}$ for the given graph $\mathbb{G}$ is its
subgraph without cycles that contains all the nodes. Figure~\ref{Fig0}
illustrates the division of a three state system into its maximal trees and
Figure~\ref{Fig1} depicts some maximal tree $\mathbb{T}$ for a six state
system. Each node $i$ specifies the realization $\mathbb{T}_{i}$ of a given
tree $\mathbb{T}$ orienting its edges in such a way that for any node $j$ they
make up the path $\mathbb{P}_{ij}$ leading from this node $j$ to the node $i$.
Then ascribing the transition rate $\left\langle i|j\right\rangle $ to the
directed edge $(i,j)$ the stationary solution of equation~(\ref{mp:1}) is
written as
\begin{equation}\label{mp:2}
    P_{i}=\frac{1}{Z}\sum_{\mathbb{T}}\prod\limits_{(kl)\in \mathbb{T}
    _{i}}\left\langle k|l\right\rangle =\frac{1}{Z}\sum_{\mathbb{T}}\exp
    \left[ -\mathcal{H}_{i}\left( \mathbb{T}\right) \right]\,,
\end{equation}
where $Z$ is the normalization constant (partition function) and the effective
energy $\mathcal{H}_{i}\left( \mathbb{T}\right) $ of the state $i$
within the tree $\mathbb{T}$ is defined as
\begin{equation}\label{mp:3}
\mathcal{H}_{i}\left( \mathbb{T}\right) =\sum_{(kl)\in \mathbb{T}_{i}}-\ln
\left[ \left\langle k|l\right\rangle \right] \,.
\end{equation}
In particular, for a given tree $\mathbb{T}$ the difference $\mathcal{H}
_{i}\left( \mathbb{T}\right) -\mathcal{H}_{j}\left( \mathbb{T}\right) $ meets
the relation
\begin{multline}\label{mp:3a}
    \mathcal{H}_{i}\left( \mathbb{T}\right) -\mathcal{H}_{j}\left( \mathbb{T}
    \right)
\\
    =\ln \left[ \prod_{(kl)\in \mathbb{P}_{ji}}\left\langle
    k|l\right\rangle \right] -\ln \left[ \prod_{(kl)\in
    \mathbb{P}_{ij}}\left\langle k|l\right\rangle \right] \,,
\end{multline}
for example, for the trees shown in Fig.~\ref{Fig1} we have
\begin{multline*}
    \mathcal{H}_{5}\left( \mathbb{T}\right) -\mathcal{H}_{1}
    \left( \mathbb{T}\right)
\\
    =\ln \left[ \left\langle 1|2\right\rangle \left\langle 2|4\right\rangle
    \left\langle 4|5\right\rangle \right] -\ln \left[ \left\langle 5|4\right\rangle
    \left\langle 4|2\right\rangle \left\langle 2|1\right\rangle \right] \,.
\end{multline*}

\begin{figure}
\begin{center}
\includegraphics{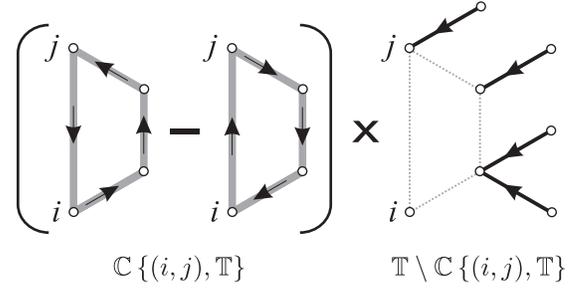}
\end{center}
\caption{Diagram visualizing an item in sum~(\protect\ref{mp:4}). Here
$\mathbb{C}\left\{ (i,j),\mathbb{T}\right\} $ is the cycle created by
connecting the nodes $j$, $i$ of the minimal tree $\mathbb{T}$, where the
forward direction of cycle tracing is given by transition from the node $j$ to
the node $i$.}
\label{Fig2}
\end{figure}

The steady state probability flux $J_{ij}$ through the edge $(ij)$ from the
node $j$ to the node $i$ is, by definition, $J_{ij}:=\left\langle
i|j\right\rangle P_{j}-\left\langle j|i\right\rangle P_{i}$. So by virtue of
(\ref{mp:2}) the equality
\begin{multline}\label{mp:4}
    J_{ij}=\frac{1}{Z}\sum_{\mathbb{T}}\left\langle i|j\right\rangle
    e^{-\mathcal{H}_{j}\left\{ \mathbb{T}\right\} }-\left\langle j|i\right\rangle
    e^{-\mathcal{H}_{i}\left\{ \mathbb{T}\right\} }
\\
    \equiv \frac{1}{Z}\sum_{\mathbb{T}}J_{\mathbb{C}\left\{ (i,j),\mathbb{T}\right\} }
    \prod\limits_{(kl)\in\mathbb{T}_{ij}}\left\langle k|l\right\rangle
\end{multline}
holds. Here $\mathbb{C}\left\{ (i,j),\mathbb{T}\right\} $ is the cycle created
via connecting the node $j$ to the node $i$ and its forward
tracing is given by the transition from $j$ to $i$, the set $\mathbb{T}_{ij}=$ $%
\mathbb{T\setminus C}\left\{ (i,j),\mathbb{T}\right\} $ is the collection of
subtrees remaining after the edges of cycle $\mathbb{C}\left\{ (i,j),\mathbb{T}%
\right\} $ having been removed from the tree $\mathbb{T}$, and
\begin{equation}
J_{\mathbb{C}\left\{ (i,j),\mathbb{T}\right\} }=\prod\limits_{(kl)\in
\mathbb{C}^{+}\left\{ (i,j),\mathbb{T}\right\} }\left\langle
k|l\right\rangle -\prod\limits_{(kl)\in \mathbb{C}^{-}\left\{ (i,j),\mathbb{T%
}\right\} }\left\langle k|l\right\rangle  \label{mp:5}
\end{equation}
is the partial probability flux through the edge $(i,j)$ related to the
cycle $\mathbb{C}\left\{ (i,j),\mathbb{T}\right\} $, where the superscripts $%
+$ and $-$ label the forward and backward directions of the cycle tracing.
We point our that the quantity $J_{\mathbb{C}\left\{ (i,j),\mathbb{T}%
\right\} }$ as well as the total probability flux $J_{ij}$ is antisymmetric
with respect to the index interchange, whereas the set $\mathbb{T}_{ij}$ and,
as a consequence, its contribution to the flux $J_{ij}$ remain the same under
this transformation. The items in sum~(\ref{mp:4}) are depicted by the diagram
in Fig.~\ref{Fig2}. The existence of nonzero probability flux under the steady
state conditions is actually the manifestation of the system being
nonequilibrium and detailed balance not holding.

\subsubsection*{Three state system. Illustrating example}

To illustrate Kirchhoff's technique described above we consider the three state
system shown in Fig.~\ref{Fig0} and imitating particle hopping in an random
medium. Without some driving field $\mathcal{E}$ the system is assumed to be
equilibrium one with the detailed balance, which is reduced to the equality
\begin{equation*}
    \left<1|2\right> \left<2|3\right>\left<3|1\right> =
    \left<1|3\right>\left<3|2\right>\left<2|1\right> \,,
\end{equation*}
implying the absence of the probability flux along the cycle 1-2-3-1. The
particle hopping between potential traps $\{u_j\}$ is described by the
transition rates
\begin{equation*}
    \left<i|j\right> = \omega e^{-u_j}\,,
\end{equation*}
where $\omega$ is some kinetic coefficient. The field $\mathcal{E}$ breaks the
detailed balance disturbing the transition rates as follows
\begin{align*}
    \left<1|2\right>_\mathcal{E} & = \left<1|2\right> e^{\mathcal{E}}, &
    \left<2|1\right>_\mathcal{E} & = \left<2|1\right> e^{-\mathcal{E}}, \\
    \left<2|3\right>_\mathcal{E} & = \left<2|3\right> e^{\mathcal{E}}, &
    \left<3|2\right>_\mathcal{E} & = \left<3|2\right> e^{-\mathcal{E}}, \\
    \left<3|1\right>_\mathcal{E} & = \left<3|1\right> e^{\mathcal{E}},  &
    \left<1|3\right>_\mathcal{E} & = \left<1|3\right> e^{-\mathcal{E}}.
\end{align*}
Then using the general expression~\eqref{mp:2} for the stationary distribution
and applying to Fig.~\ref{Fig0} showing the corresponding collection of maximal
trees we get the expression for the partition function
\begin{equation*}
    Z(\mathcal{E})   = \omega^2 e^{-u_1-u_2-u_3}
                \left(e^{u_1}+ e^{u_2}+e^{u_3}\right)
                \left[1+2\cosh( 2\mathcal{E})\right].
\end{equation*}
Then by virtue of \eqref{mp:4} the probability flux along any edge of this
system is
\begin{equation*}
    J = \frac{2\omega\sinh\mathcal{E}}{(e^{u_1}+e^{u_3}+e^{u_3})}\,.
\end{equation*}

Below the effective energy $\mathcal{H}_i(\mathbb{T})$ for the state $i$ within
a given tree $\mathbb{T}$ has been introduced by expression~\eqref{mp:3}. For
the three state system under consideration applying to Fig.~\ref{Fig0} we can
write, for example,
\begin{align*}
    \mathcal{H}_2(\mathbb{T}_1) - \mathcal{H}_1(\mathbb{T}_1) & = 2\mathcal{E}
    - (u_2 -u_1)\,,
\\
    \mathcal{H}_3(\mathbb{T}_1) - \mathcal{H}_1(\mathbb{T}_1) & = 4\mathcal{E}
    - (u_3 -u_1)\,,
\\
    \mathcal{H}_2(\mathbb{T}_2) - \mathcal{H}_1(\mathbb{T}_2) & = 2\mathcal{E}
    - (u_2 -u_1)\,,
\\
    \mathcal{H}_3(\mathbb{T}_2) - \mathcal{H}_1(\mathbb{T}_2) & = -2\mathcal{E}
    - (u_3 -u_1)\,.
\end{align*}
Whence it follows that the effective energies $\mathcal{H}_i(\mathbb{T}_1)$,
$\mathcal{H}_i(\mathbb{T}_2)$ of the trees $\mathbb{T}_1$, $\mathbb{T}_2$
cannot be represented in terms of one energy $H^{ss}_i$ multiplied by some
individual cofactors $\beta_1$, $\beta_2$, i.e. there is no function $H^{ss}_i$
such that $\mathcal{H}_i(\mathbb{T}_1) \mapsto \beta_1 H^{ss}_i$ and
$\mathcal{H}_i(\mathbb{T}_2) \mapsto \beta_2 H^{ss}_i$. Indeed, otherwise, the
relations $\beta_1=\beta_2$ and $\beta_1 \neq\beta_2$ have to hold
simultaneously.

\subsection*{Equivalence classes of maximal trees. Channels}

\begin{figure}
\begin{center}
\includegraphics{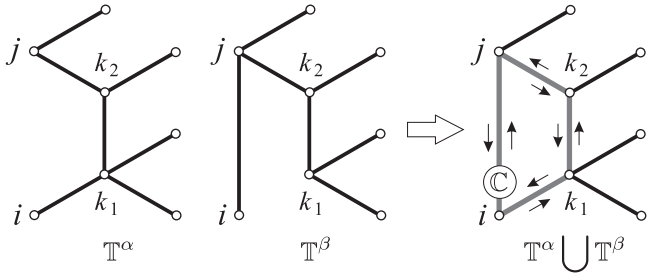}
\end{center}
\caption{Illustration of the cycle formation by the superposition of maximal
trees.}
\label{Fig3}
\end{figure}

At the next step a special case of Markovian systems is analyzed, where there
are collections of many maximal trees within which the detailed balance holds
individually. Namely, let us consider some two maximal trees $\mathbb{T
}^{\alpha }$, $\mathbb{T}^{\beta }$ , and the union of their edges $\mathbb{T
}^{\alpha }\bigcup \mathbb{T}^{\beta }$ referred below as to the tree
superposition. The two trees are said to be equivalent, $\mathbb{T}^{\alpha
}\sim \mathbb{T}^{\beta }$, if their superposition does not contain any cycle
$\mathbb{C}$ of edges with nonzero probability flux $J_{\mathbb{C}}\neq 0$,
which is illustrated in Fig.~\ref{Fig3}. Applying, for example, to the cycle
$\mathbb{C}$ shown in Fig.~\ref{Fig3} the condition of zero probability flux
reads
\begin{equation*}
\left\langle i|j\right\rangle \left\langle j|k_{2}\right\rangle \left\langle
k_{2}|k_{1}\right\rangle \left\langle k_{1}|i\right\rangle =\left\langle
i|k_{1}\right\rangle \left\langle k_{1}|k_{2}\right\rangle \left\langle
k_{2}|j\right\rangle \left\langle j|i\right\rangle
\end{equation*}
thus
\begin{equation*}
\frac{\left\langle j|k_{2}\right\rangle \left\langle k_{2}|k_{1}\right\rangle
\left\langle k_{1}|i\right\rangle }{\left\langle i|k_{1}\right\rangle
\left\langle k_{1}|k_{2}\right\rangle \left\langle k_{2}|j\right\rangle
}=\frac{\left\langle j|i\right\rangle }{\left\langle i|j\right\rangle }
\end{equation*}
whence, by virtue of (\ref{mp:3a}), we get the identity
\begin{equation*}
\mathcal{H}_{i}\left( \mathbb{T}^{\alpha }\right) -\mathcal{H}_{j}\left(
\mathbb{T}^{\alpha }\right) =\mathcal{H}_{i}\left( \mathbb{T}^{\beta }\right)
-\mathcal{H}_{j}\left( \mathbb{T}^{\beta }\right) \,.
\end{equation*}
This example demonstrates the fact that for equivalent trees, e.g., $\mathbb{%
T}^{\alpha }$and $\mathbb{T}^{\beta }$ their effective energies differ in
constant value only, with latter statement being actually an equipollent
definition of the tree equivalence,
\begin{equation}
\mathbb{T}^{\alpha }\sim \mathbb{T}^{\beta }\Longleftrightarrow \mathcal{H}%
_{i}\left( \mathbb{T}^{\alpha }\right) -\mathcal{H}_{i}\left( \mathbb{T}%
^{\beta }\right) =\text{const}\,.  \label{mp:6}
\end{equation}
Thereby the introduced relationship between the maximal trees really form an
equivalence relation because the condition $\mathbb{T}^{\alpha }\sim \mathbb{%
T}^{\beta }$ and $\mathbb{T}^{\alpha }\sim \mathbb{T}^{\gamma }$ gives rise to
$\mathbb{T}^{\beta }\sim \mathbb{T}^{\gamma }$. Therefore the collection of all
the maximal trees $\left\{ \mathbb{T}\right\} $ of the state graph $\mathbb{G}$
can be divided in the classes $\left\{ \mathbb{K}\right\} $ of equivalent trees
that will be called channels. The superposition of all the maximal trees
belonging to one channel $\mathbb{K}$ i.e. its implementation as a subgraph of
graph $\mathbb{G}$ will be also referred to as just the channel $\mathbb{K}$.

\begin{figure}
\begin{center}
\includegraphics{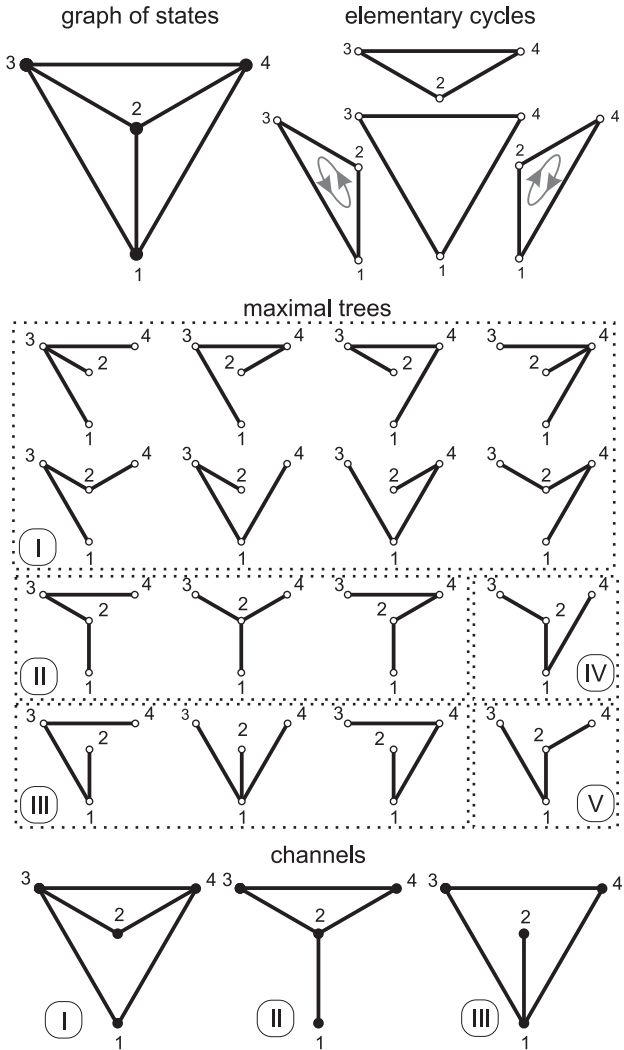}
\end{center}
\caption{An example of four state system and its division into channels I--V.
In addition the elementary cycles with zero and nonzero probability flux are
shown.}
\label{Fig4}
\end{figure}

The fact that the notion of channels does have some meaning is justified in
Fig.~\ref{Fig4} exhibiting a case where there are channels contaning more then
one maximal tree and not coinciding with the system as a whole. Namely, this
figure depicts a four state system for which there are two elementary (three
state) cycles with zero probability flux (cycles 1-3-4-1 and 2-3-4-2) and two
ones with nonzero flux (cycles 1-3-2-1 and 1-2-4-1). These cycles, indeed, can
have such properties, which requires some comments. The matter is that the
fluxes $J_{\text{1-3-2-1}}$ and $J_{\text{1-2-4-1}}$ are not independent
because the compound cycle 1-3-2-4-1 can be obtained unifying either the cycles
1-3-2-1 and 1-2-4-1 or the cycles 1-3-4-1 and 2-3-4-2. The unification of
cycles with zero probability flux inevitably gives rise to the compound cycle
with zero flux too. So the fluxes $J_{\text{1-3-2-1}}$ and $J_{\text{1-2-4-1}}$
should also meet the condition of zero probability flux for the cycle
1-3-2-4-1. The detailed analysis of the relationship between the probability
fluxes of elementary and compound cycles is beyond the scope of the present
paper. Here we note only that the adopted flux pattern can be implemented for
the given graph when all its edges are symmetrical with respect to the
transition rate, $\left\langle
i|j\right\rangle =\left\langle j|i\right\rangle $, except for the edge $%
\left( 1,2\right) $, where $\left\langle 1|2\right\rangle \neq \left\langle
2|1\right\rangle $. The induced partition of the maximal tree collection
into the channels and their implementation in graph form are shown in Fig.~%
\ref{Fig4}.

\subsection*{Channel superstatistics of the steady state distribution}

In order to describe the properties of a maximal tree $\mathbb{T}$ within its
channel $\mathbb{K}$ containing $N_{\mathbb{K}}$ trees, first, the effective
energy $\mathcal{H}_{i}\left\{ \mathbb{K}\right\} $ averaged over the given
channel
\begin{equation}
H_{\mathbb{K}}(i)=\frac{1}{N_{\mathbb{K}}}\sum_{\mathbb{T}^{\prime }\in
\mathbb{K}}\mathcal{H}_{i}\left( \mathbb{T}^{\prime }\right)  \label{mp:7}
\end{equation}
is introduced. The quantity $H_{i}\left( \mathbb{K}\right) $ will be called the
effective energy of the state $i$ within the channel $\mathbb{K}$. Then, using
the constructed function the effective energy $\mathcal{H}_{i}\left(
\mathbb{T}\right) $ of the state $i$ within the tree $\mathbb{T}$ in rewritten
as
\begin{equation}
\mathcal{H}_{i}\left( \mathbb{T}\right) =H_{\mathbb{K}}(i)+U\left( \mathbb{T}%
|\mathbb{K}\right) \,,  \label{mp:8}
\end{equation}
where the value
\begin{equation}
U\left( \mathbb{T}|\mathbb{K}\right) :=\frac{1}{N_{\mathbb{K}}}\sum_{\mathbb{%
T}^{\prime }\in \mathbb{K}}\left[ \mathcal{H}_{i}\left( \mathbb{T}^{\prime
}\right) -\mathcal{H}_{i}\left( \mathbb{T}\right) \right]  \label{mp:9}
\end{equation}
is regarded as the effective energy of the tree $\mathbb{T}$ within the channel
$\mathbb{K}$ because due to property~(\ref{mp:6}) it does not depend on the
state $i$ and, thus, characterizes the tree $\mathbb{T}$ as a whole. Naturally,
the mean value of $U\left( \mathbb{T}|\mathbb{K}\right)$ is equal to zero,
\begin{equation}
V_{\mathbb{K}}=\frac{1}{N_{\mathbb{K}}}\sum_{\mathbb{T}\in \mathbb{K}}U\left(
\mathbb{T}|\mathbb{K}\right) =0\,.  \label{mp:9a}
\end{equation}

The introduced quantities permit us to rewrite expression~(\ref{mp:2}) for the
steady state distribution of the given Markovian system as a sum running over
the channels
\begin{align}
\nonumber
    P_{i}& =\frac{1}{Z}\sum_{\mathbb{K}}N_{\mathbb{K}}\text{e}^{-F_{\mathbb{K}}}
    \exp \left\{ -H_{\mathbb{K}}(i)\right\}
\\
\label{result:1}
    {}& =\frac{1}{Z}\sum_{\mathbb{K}}w(\mathbb{K})\exp
    \left\{ -H_{\mathbb{K}}(i)\right\} \,.
\end{align}
Here the quantity $F_{\mathbb{K}}$ has appeared in formula~(\ref{result:1})
via the sum over all the maximal trees $\left\{ \mathbb{T}\right\} _{\mathbb{%
K}}$ composing the channel $\mathbb{K}$
\begin{equation}
\exp \left( -F_{\mathbb{K}}\right) :=\frac{1}{N_{\mathbb{K}}}\sum_{\mathbb{T}%
\in \mathbb{K}}\exp \left\{ -U\left( \mathbb{T}|\mathbb{K}\right) \right\}
\label{add:2}
\end{equation}
and specifies the statistical weight of channel $\mathbb{K}$
\begin{equation*}
w(\mathbb{K})=N_{\mathbb{K}}e^{-F_{\mathbb{K}}}\,.
\end{equation*}
In order to write the desired expression for $F_{\mathbb{K}}$ we make use of
the expansion
\begin{equation*}
e^{-U}=1-U+\varphi (U)\,,\ \text{where}\ \varphi (U)>0\ \text{for}\ U\neq 0\,.
\end{equation*}
Whence, by virtue of (\ref{mp:9a}),
\begin{equation} \label{add:3}
    \exp \left( -F_{\mathbb{K}}\right) =1+\frac{1}{N_{\mathbb{K}}}
    \sum_{\mathbb{T}\in \mathbb{K}}\varphi
    \left[U\left( \mathbb{T}|\mathbb{K}\right) \right]
    \overset{\text{def}}{=}\exp \left\{ \theta _{\mathbb{K}}
    S_{\mathbb{K}}\right\}\,,
\end{equation}
where $S_{K}:=\ln N_{\mathbb{K}}$ has the meaning of channel entropy. Due to
the zero value of the mean channel energy $V_{\mathbb{K}}=0$ the quantity
$F_{\mathbb{K}}$ can be rewritten as $F_{\mathbb{K}}=V_{\mathbb{K}}-\theta
_{\mathbb{K}}S_{\mathbb{K}}$ and keeping in mind the standard notions of
statistical physics it will be called the free energy of channel $\mathbb{K}$.
In particular, the quantity
\begin{equation} \label{add:4}
    \theta _{\mathbb{K}}=\frac{1}{\ln N_{\mathbb{K}}}\ln \left\{ 1+\frac{1}
    {N_{\mathbb{K}}}\sum_{\mathbb{T}\in \mathbb{K}}\varphi
    \left[ U\left( \mathbb{T}|\mathbb{K}\right) \right] \right\}
\end{equation}
is an order parameter of the channel structure; when all the trees of a channel
$\mathbb{K}$ have the same energy $U\left( \mathbb{T}|\mathbb{K} \right)
=U\left( \mathbb{K}\right) $ the value $\theta =0$ and the wider the
distribution of the tree energies, the large the value $\theta _{\mathbb{K}}$.

Finalizing this section we rewrite formula~\eqref{result:1} as
\begin{equation}\label{abzas}
    P_{i} =
    \frac{1}{Z}\sum_{\mathbb{K}}N_\mathbb{K}
    \exp\left\{\theta_\mathbb{K} S_\mathbb{K} -H_{\mathbb{K}}(i)\right\} \,.
\end{equation}
Expression~\eqref{abzas} is the desired superstatistics representation of the
steady state distribution for a nonequilibrium Markovian system. It should be
noted that the obtained effective energy $H_{\mathbb{K}}(i)$ of the system
states $\left\{ i\right\} $ within the channel $\mathbb{K}$ can depend on many
parameters of this channel and is reduced to some fixed function $H(i)$
multiplied by an inverse channel \textquotedblleft
temperature\textquotedblright , $H_{\mathbb{K}}(i)=\beta _{\mathbb{K}}H(i)$, in
special cases only. This could take place if, for example, the distribution of
the tree energies within one channel and the distribution of the system states
are caused by the same mechanism. In this case it might be expected that the
channel temperature will be specified by the order parameter $\theta
_{\mathbb{K}}$, i.e. $1/\beta _{\mathbb{K}}=\theta _{\mathbb{K}}$.

\begin{figure}
\begin{center}
\includegraphics[width=0.9\columnwidth]{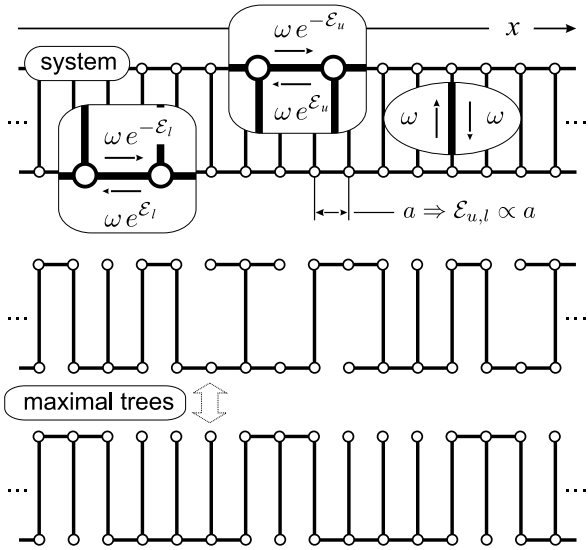}
\caption{Example of a system where at least some channels can be characterized
by the effective energies of the states differing only by constant prefactors.}
\label{Fig00}
\end{center}
\end{figure}

Figure~\ref{Fig00} visualizes an example of systems where differen channels, at
least some ones, can be ascribed with the effective energies for the system
states that differ only by some constant prefactors. It is a stair case system
with the transition rates shown in Fig.~\ref{Fig00}, where the lower two
diagrams depict equivalent maximal trees. Rigorously speaking, these trees
become equivalent only on scales $\delta x \gg a$ or, what is actually the
same, in the limit $a\to 0$. Indeed, at the first approximation the energy of a
state $i$ located near the point $x$ along the system (Fig.~\ref{Fig00}) can be
written as
\begin{equation*}
  H_{\mathbb{K}}(x) = 2\big(q_u\mathcal{E}_u+q_l\mathcal{E}_l\big)
  \frac{x}{a} + \text{const}\,,
\end{equation*}
where $q_{u,l}$ is the relative portion of edges on the upper and lower
branches belonging to the trees of a given channel, $q_{u}+q_{l} =1$. In some
sense the value $q_{u}$ (or $q_{l}$, what is the same) is the main
characteristics of such a channel. It is assumed to be distributed rather
uniformly along the system because exactly the uniform distribution matches the
maximal number of the channel realizations in the maximal trees. Naturally,
there are channels in this system where the portion $q_{u}$ and, thus, also
$q_{l}$ can depend on the coordinate $x$ and, thereby, their effective energies
are not reduced to the presented form. However, their contribution to the
steady state distribution could be rather small.

\subsection*{Channel superstatistics of the steady state probability flux}

\begin{figure}
\begin{center}
\includegraphics{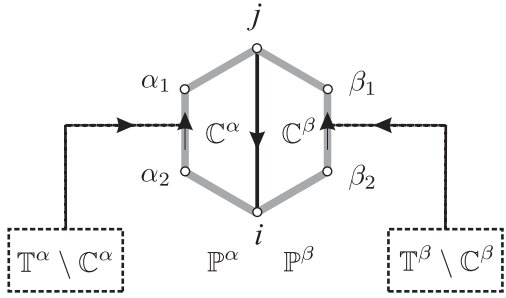}
\end{center}
\caption{Two cycles $\mathbb{C}^{\protect%
\alpha }$ and $\mathbb{C}^{\protect\beta }$ created via the connection of the
nodes $i$ and $j$ on the maximal trees $\mathbb{T}^{\protect\alpha }$ and
$\mathbb{T}^{\protect\beta }$ belonging to one channel $\mathbb{K}$. Arrows
specify the forward direction of the cycle tracing.}
\label{Fig5}
\end{figure}

The steady state probability flux $\{J_{ij}\}$ through the graph edges can be
also reduced to a sum over the channels. In order to do this we make use of
formula~(\ref{mp:4}). Let us consider an edge $\left( i,j\right) $ and two
trees $\mathbb{T}^{\alpha },$ $\mathbb{T}^{\beta }\in \mathbb{K}$ belonging to
one channel $\mathbb{K}$. The given edge $\left( i,j\right) $ is assumed
beforehand not to belong to the channel $\mathbb{K}$ because otherwise all its
maximal trees do not contribute to the flux $J_{ij}$. The edge $\left(
i,j\right) $ with the edges of the trees $\mathbb{T}^{\alpha }$ and
$\mathbb{T}^{\beta }$ form two cycles $\mathbb{C}^{\alpha }=\left( i,j\right)
\bigcup \mathbb{P}_{i,j}^{\alpha }$ and $\mathbb{C}^{\beta
}=\left( i,j\right) \bigcup \mathbb{P}_{i,j}^{\beta }$, where $\mathbb{P}%
_{i,j}^{\alpha }$, $\mathbb{P}_{i,j}^{\beta }$ are paths on the trees $%
\mathbb{T}^{\alpha }$, $\mathbb{T}^{\beta }$, respectively, connecting the
nodes $i$ and $j$ (Fig.~\ref{Fig5}). The paths $\mathbb{P}_{i,j}^{\alpha }$,
$\mathbb{P}_{i,j}^{\beta }$ can coincide with each other partly or even
completely. Since the two trees belong to one channel the probability flux
along the composite cycle $\mathbb{C}^{\alpha \beta }=\mathbb{C}^{\alpha
}\bigcup \mathbb{C}^{\beta }= \mathbb{P}_{i,j}^{\alpha }\bigcup
\mathbb{P}_{i,j}^{\beta }$ is equal to zero. The latter condition is
implemented by the equality
\begin{equation}
\prod\limits_{(kl)\in \mathbb{P}^{\alpha +}}\left\langle k|l\right\rangle
\prod\limits_{(kl)\in \mathbb{P}^{\beta -}}\left\langle k|l\right\rangle
=\prod\limits_{(kl)\in \mathbb{P}^{\alpha -}}\left\langle k|l\right\rangle
\prod\limits_{(kl)\in \mathbb{P}^{\beta +}}\left\langle k|l\right\rangle \,,
\label{flux:1}
\end{equation}
in particular, for the case shown in Fig.~\ref{Fig5}
\begin{multline*}
    \left\langle j|\alpha _{1}\right\rangle \left\langle \alpha _{1}|\alpha
    _{2}\right\rangle \left\langle \alpha _{2}|i\right\rangle \left\langle i|\beta
    _{2}\right\rangle \left\langle \beta _{2}|\beta _{1}\right\rangle \left\langle
    \beta _{1}|j\right\rangle
\\
    =\left\langle j|\beta _{1}\right\rangle \left\langle
    \beta _{1}|\beta _{2}\right\rangle \left\langle \beta _{2}|i\right\rangle
    \left\langle i|\alpha _{2}\right\rangle \left\langle \alpha _{2}|\alpha
    _{1}\right\rangle \left\langle \alpha _{1}|j\right\rangle \,.
\end{multline*}
Let us introduce the intensities $R_{ij}^{\alpha }$, $R_{ij}^{\beta }$, and the
asymmetry $\mathcal{A}_{ij}\left( \mathbb{K}\right) $ of the transitions along
the paths $\mathbb{P}_{i,j}^{\alpha }$, $\mathbb{P}_{i,j}^{\beta }$
via the expressions
\begin{eqnarray}
\label{flux:2}
    R_{ij}^{\alpha ,\beta } &=&\left[ \prod\limits_{(kl)\in \mathbb{P}^{\alpha
    ,\beta +}}\left\langle k|l\right\rangle \prod\limits_{(kl)\in
    \mathbb{P}^{\alpha ,\beta -}}\left\langle k|l\right\rangle \right] ^{1/2},
\\
\nonumber
    e^{\mathcal{A}_{ij}\left( \mathbb{K}\right)}  &=&\left[
    \prod\limits_{(kl)\in \mathbb{P}^{\alpha +}}\left\langle k|l\right\rangle
    \right] ^{1/2}\cdot \left[ \prod\limits_{(kl)\in \mathbb{P}^{\alpha
    -}}\left\langle k|l\right\rangle \right] ^{-1/2}
\\
\label{flux:3}
    &=&\left[ \prod\limits_{(kl)\in \mathbb{P}^{\beta +}}\left\langle
    k|l\right\rangle \right] ^{1/2}\cdot \left[ \prod\limits_{(kl)\in
    \mathbb{P}^{\beta -}}\left\langle k|l\right\rangle \right] ^{-1/2}
\end{eqnarray}
with the value $\mathcal{A}_{ij}\left( \mathbb{K}\right) $ being the same for
both the paths $\mathbb{P}_{i,j}^{\alpha }$, $\mathbb{P}_{i,j}^{\beta }$ due to
(\ref{flux:1}). In other words, the quantity $\mathcal{A}_{ij}\left(
\mathbb{K}\right) $ is the characteristics of  the edge $\left( i,j\right) $
with respect to the channel $\mathbb{K}$ rather than to its trees individually.
It should be pointed out that the quantities $R_{ij}^{\alpha } $ are
symmetrical whereas $\mathcal{A}_{ij}\left( \mathbb{K}\right) $ is asymmetrical
with respect to the index interchange, i.e., $R_{ij}^{\gamma
}=R_{ji}^{\gamma }$ and $\mathcal{A}_{ij}\left( \mathbb{K}\right) =-\mathcal{%
A}_{ji}\left( \mathbb{K}\right) $. In addition, let the quantities $%
P_{ij}^{\alpha }$ and $P_{ij}^{\beta }$ stand for the contributions of the
subtree collections $\mathbb{T}^{\alpha }\setminus \mathbb{C}^{\alpha }$ and
$\mathbb{T}^{\beta }\setminus \mathbb{C}^{\beta }$ (Fig.~\ref{Fig2}) to the
tree weights $\exp \left\{ -\mathcal{H}_{i}\left( \mathbb{T}^{\alpha
}\right) \right\} $ and $\exp \left\{ -\mathcal{H}_{i}\left( \mathbb{T}%
^{\beta }\right) \right\} $, respectively. As noted above the quantities $%
P_{ij}^{\gamma }$ are symmetrical within the index interchange. Applying to
definition~(\ref{mp:3}) of the effective energy $\mathcal{H}_{i}\left(
\mathbb{T}^{\gamma }\right) $ we write ($\gamma =\alpha ,\beta $)
\begin{eqnarray*}
\exp \left\{ -\mathcal{H}_{i}\left( \mathbb{T}^{\gamma }\right) \right\}
&=&R_{ij}^{\gamma }P_{ij}^{\gamma }\exp \left\{ -\mathcal{A}_{ij}\left(
\mathbb{K}\right) \right\} \,, \\
\exp \left\{ -\mathcal{H}_{j}\left( \mathbb{T}^{\gamma }\right) \right\}
&=&R_{ij}^{\gamma }P_{ij}^{\gamma }\exp \left\{ \mathcal{A}_{ij}\left(
\mathbb{K}\right) \right\} \,,
\end{eqnarray*}
whence taking also into account relationship~(\ref{mp:8}) between the
effective energies of the system state $i$ within the tree $\mathbb{T}%
^{\gamma }$ and the corresponding channel $\mathbb{K}$ we get
\begin{eqnarray*}
R_{ij}^{\gamma }P_{ij}^{\gamma } &=&\exp \left\{ -\frac{1}{2}\left[ \mathcal{%
H}_{i}\left( \mathbb{T}^{\gamma }\right) +\mathcal{H}_{j}\left( \mathbb{T}%
^{\gamma }\right) \right] \right\}  \\
&=&\exp \left\{ -\frac{1}{2}\left[ H_{\mathbb{K}}(i)+H_{\mathbb{K}}(j)\right]
-U\left( \mathbb{T}^{\gamma }|\mathbb{K}\right) \right\} \,.
\end{eqnarray*}

In these terms the partial probability flux~(see (\ref{mp:4}) and (\ref{Fig5}%
)) for the two trees $\mathbb{T}^{\alpha }$, $\mathbb{T}^{\beta }$
individually becomes ($\gamma =\alpha ,\beta $)
\begin{multline}\label{flux:10}
    J_{ij}\left( \mathbb{T}^{\gamma }\right)  = J_{\mathbb{C}\left\{ (i,j),
    \mathbb{T}^{\gamma }\right\} }\prod\limits_{(kl)\in
    \mathbb{T}_{ij}^{\gamma}}\left\langle k|l\right\rangle
\\
    = \left[ \left\langle i|j\right\rangle \exp \left\{ \mathcal{A}_{ij}\left(
    \mathbb{K}\right) \right\} -\left\langle j|i\right\rangle \exp
    \left\{ -\mathcal{A}_{ij}\left( \mathbb{K}\right) \right\} \right] R_{ij}^{\gamma
    }P_{ij}^{\gamma }
\\
    = \mathcal{E}_{ij}\left( \mathbb{K}\right) \exp \left\{ -\frac{1}{2}\left[
    H_{\mathbb{K}}(i)+H_{\mathbb{K}}(j)\right] -U\left( \mathbb{T}^{\gamma }|\mathbb{K}\right)
    \right\} \,,
\end{multline}
where the quantity $\mathcal{E}_{ij}\left( \mathbb{K}\right) $ introduced
by the expression
\begin{equation}
\mathcal{E}_{ij}\left( \mathbb{K}\right) :=\left[ \left\langle i|j\right\rangle
\exp \left\{ \mathcal{A}_{ij}\left( \mathbb{K}\right) \right\} -\left\langle
j|i\right\rangle \exp \left\{ -\mathcal{A}_{ij}\left( \mathbb{K}\right)
\right\} \right]   \label{flux:11}
\end{equation}
is ascribed directly to the edge $\left( i,j\right) $ with respect to the
channel $\mathbb{K}$. It characterizes the contribution of the channel
\thinspace $\mathbb{K}$ to the stationary probability flux through the edge $%
\left( i,j\right) $, namely, by virtue of (\ref{mp:4}), (\ref{add:2}), and (%
\ref{flux:10})
\begin{equation}
J_{ij}=\frac{1}{Z}\sum_{\mathbb{T}}\mathcal{E}_{ij}\left( \mathbb{K}\right)
N_{\mathbb{K}}e^{-F_{\mathbb{K}}}\exp \left\{ -\frac{1}{2}\left[ H_{\mathbb{K%
}}(i)+H_{\mathbb{K}}(j)\right] \right\} \,.  \label{flux:12}
\end{equation}
Expression~(\ref{flux:12}) is the desired superstatistics representation of the
stationary flux of probability. In some sense the nonequilibrium properties of
the Markovian system under consideration are behind the driven forces $\left\{
\mathcal{E}_{ij}\left( \mathbb{K}\right) \right\} $ induced by channel
$\mathbb{K}$. In particular, if the edge $\left( i,j\right) $ belongs to the
channel $\mathbb{K}$ then $\mathcal{E}_{ij}\left( \mathbb{K}\right) =0$.

\section{Conclusion}

In conclusion we have derived a general form of the stationary probability
distribution for Markovian systems. This has allowed us to obtain a novel
interpretation of probability distributions used in the framework of
superstatistics, or more rigorously, generalized superstatistics. This notion
has been successful in reproducing the probability distribution of many systems
ranging from turbulence to economics. However, the justification of the form of
the probability distribution as a weighted sum over equilibrium distributions
up to now is based on purely phenomenological arguments. No fundamental
derivation from the basic principles has been given.

In the present paper we have been able to derive an analogy to the
superstatistics approximation with respect to the probability distribution as
well as the stationary probability fluxes. This means that superstatistical
representations can also be formulated for fluxes, which has not been discussed
up to now. Our treatment is based on Kirchhoff's diagram technique applied to
the master equation for Markovian processes. This techniques has been known for
many years. We have included an assumption on the properties of the system
graphs. As a main point we have focused on systems possessing a wide collection
of cycles with vanishing fluxes. In this case it is possible to construct
equivalence classes of Kirchhoff's maximal trees called channels, for which
detailed balance holds individually. The nonequilibrium properties initially
attributed to cycles can be assigned to individual edges, i.e. to individual
transitions between a pair of states within one channel.

It is worthwhile to underline the fact that in the present paper we actually
have formulated an original approach to describing steady states of
nonequilibrium systems. In its spirit it is rather similar to the widely used
notion of superstatistics but, nevertheless, differs from at the basics. The
key point of the developed approach is the representation of a nonequilibrium
system as a superposition of its equilibrium subsystems (channels) with local
detailed balance rather then a formal expansion of the steady state
distribution into the sum over terms of the Boltzmann type. Our considerations
shed light on the network structures of complex systems and may help to
understand systems in flux equilibrium like turbulent flows, traffic flows and
economic systems. Beside, the introduced notion of channels enables one to pose
a question as to whether it is possible to describe the dynamics of Markovian
systems as transient processes in the channels individually and their
interaction with one another.

\begin{acknowledgments}
The authors appreciate the financial support of the SFB 458 and the University
of M\"unster as well as the partial support of DFG project MA 1508/8-1 and RFBR
grants 06-01-04005, 05-01-00723, and 05-07-90248.
\end{acknowledgments}

\end{document}